\documentclass[12pt]{article}

\usepackage{caption}
\usepackage{color}

\usepackage{epsf}
\usepackage{pazha}

\usepackage{graphicx}
\usepackage{wrapfig}

\usepackage{amsmath}
\usepackage{amssymb}

\tightenlines

\def\arcdeg{\hbox{$^\circ$}}

\def\arcsec{\hbox{$^{\prime\prime}$}}

\newcommand {\be}{\begin {equation}}
\newcommand {\ee}{\end {equation}}

\def\*{$^{*}$}

\sloppypar

\begin{document}

{\it Will be published in Astronomy Letters, 2017, v.43, N4}

\title{\bf OBSERVATIONS OF THE X-RAY PULSAR LMC\,X-4 WITH \textit{NuSTAR}:
LIMIT ON THE MAGNETIC FIELD AND TOMOGRAPHY OF THE SYSTEM IN THE
FLUORESCENT IRON LINE}

\author{\bf \hspace{-1.3cm}\copyright\, 2016 \ \
A.E.Shtykovsky\affilmark{1*}, A.A.Lutovinov\affilmark{1}, V.A.Arefiev\affilmark{1}, S.V.Molkov\affilmark{1}, S.S.Tsygankov\affilmark{2}}

\affil{
{$^1$ \it Space Research Institute of the Russian Academy of Sciences, Moscow, Russia}\\
{$^2$ \it Tuorla Observatory, University of Turku, Turku, Finland}}

\vspace{2mm}

\sloppypar
\vspace{2mm}
\noindent We present results of the spectral and timing analysis of the X-ray
pulsar LMC\,X-4 with the {\it NuSTAR} observatory in the broad energy range
3-79 keV. Along with the detailed analysis of the averaged source spectrum,
the high-precision pulse phased-resolved spectra were obtained for the first
time. It has been shown that the comptonization model gives the best
approximation of the obtained spectra. The evolution of its parameters was
traced depending on the pulse phase as well. The search for the possible
cyclotron absorption line was performed for all energy spectra in the
5-55\,keV energy range. The obtained upper limit for the depth of the
cyclotron absorption line $\tau \sim 0.15$ ($3\sigma$) indicates no cyclotron
absorption line in this energy range, which provides an estimate of the
magnitude of the magnetic field on the surface of the neutron star: $B < 3
\times 10^{11} \text{ G}$ or $B > 6.5 \times 10^{12} \text{ G}$. The latter
one is agree with the estimate of the magnetic field obtained from the
analysis of the power spectrum of the pulsar: $B \sim 3 \times 10^{13} \text{
G}$. Based on results of the pulse phase-resolved spectroscopy we revealed a
delay between maxima of the emission and the equivalent width of the
fluorescent iron line. This delay can be apparently associated with the
travel time of photons between the emitting regions in the vicinity of the
neutron star and the relatively cold regions where this emission is reflected
(presumably, at the inflowing stream or at the place of an interaction of the
stream and the outer edge of the accretion disk).


\vfill
\noindent\rule{8cm}{1pt}\\
{$^*$ E-mail $<$sht.job@gmail.com$>$}

\clearpage

\section*{INTRODUCTION }

\noindent The X-ray pulsar LMC\,X-4, discovered by the {\it Uhuru} observatory
(Giacconi et~al., 1972), is a high-mass binary system, located in the Large
Magellanic Cloud (estimated distance $d = 50 \text{ kpc}$). The system
consists of a neutron star with the mass of $M_{\ast} \simeq 1.57M_{\sun}$,
where $M_{\sun}$ -- is the Solar mass, and an optical counterpart -- the
O8III spectral class star with the mass of $\sim 18 M_{\sun}$ (see Falanga et
al. 2015, and references therein). The system orbital period is about of
$P_{\rm orb} \simeq 1.4 \text{ days}$. In the early papers on LMC\,X-4 Lee et
al. (1978) and White (1978) have shown an eclipsing nature of the pulsar in
X-rays, which is connected with a significant inclination of the system
to the observer.

Besides the spin period and the orbital motion, the system exhibits a so-called
superorbital variability (Lang et al., 1981), during which an intensity of
the source changes for more than $\sim 50$ times during a characteristic
period of $P_{\rm sup} \simeq 30.5 \text{ days}$. Presence of the
superorbital period is believed to be due to the obscuration of the direct
radiation from the neutron star by the precessing accretion disk (Lang et al.
1981; Heemskerk, van Paradijs 1989; Neilsen et al. 2009).

The most accurate ephemeris for the orbital parameters, including the decay
of the orbital period and the superorbital variability, were obtained in
recent papers by Falanga et al. (2015) and Molkov et al. (2015). The maximum
value of the persistent luminosity of the source in X-ray energy range is
about $L_{\rm x} \simeq (3-4) \times 10^{38} \text{ erg s}^{-1}$ (La Barbera
et al. 2001; Tsygankov, Lutovinov 2005; Grebenev et al. 2013), which is close
to the Eddington limit of the luminosity of an accreting neutron star (note,
that in a presence of a strong magnetic field this limit may be much higher,
see Mushtukov et al. 2015).

In addition to the periodic changes of the intensity of the binary system,
aperiodic series of short flares are observed in X-rays. During such events
the X-ray intensity of the source can be increased ten times compared to the
maximum persistent flux (see, e.g., Epstein et al. 1977; Levine et al. 2000;
Moon et al. 2001).

Pulsations with the period of $P_{\rm spin}\simeq13.5 \text{ sec}$ were
discovered in X-rays using the {\it SAS-3} observatory data during such
flares (Kelley et al. 1983). Measurements of the spin period, carried out
over several decades have shown that it doesn't remain constant over the time
and varies near the average value in a nearly periodical manner (Molkov et
al. 2016). In the same paper, several mechanisms have been considered,
capable of providing the observed behavior of the pulsation period -- from
the presence of a third body in the system to the switching between
different states of the pulsar magnetosphere.

If we assume that the pulsar is in the equilibrium state due to the balance
of accelerating and decelerating torques, the combination of the high
luminosity of LMC~X-4 and its relatively small period leads to a relatively
high estimate of the magnetic field on the surface of the neutron star $B
\geq 10^{13} \text{ G}$ (Ghosh, Lamb 1979; Naranan et al. 1985).

In X-ray pulsars one of the most direct and reliable way to determine the
magnetic field is the registration of the cyclotron absorption lines in their
energy spectra (the most full to date list of objects having such features is
presented in the review by Walter et al. 2015). It is important to note, that
sometimes cyclotron absorption lines, poorly registered or even not
registered in average spectra of pulsars, were found in the phase-resolved
spectra. The power-law with the high-energy cutoff model (White et al. 1983;
Woo et al. 1996) as well as a comptonization model (La Barbera et al. 2001)
are usually used to describe the spectrum of LMC~X-4. The $\rm K_{\alpha}$
iron line at the energy of $E_{\rm Fe} \simeq6.4 \text{ keV}$ is also
detected in the source spectrum (Nagase 1989; Levine et al. 1991). The value
of the absorption column in a direction towards the source is of $N_{\rm H}
\sim 5.74 \times 10^{20}\,\text{atoms cm}^{-2}$ (Hickox 2004) that close to
the value of the galactic absorption, indicating absence of a significant
intrinsic absorption in the binary system. Previous searches for
the cyclotron feature in the spectrum of LMC~X-4 in the energy range up to
100 keV based on the {\it Ginga, RXTE} and {\it INTEGRAL} observatory data
have not given any positive result (see, e.g., Levine et al. 1991; Woo et
al. 1996; Tsygankov, Lutovinov 2005). The only mention of a possible
presence of the cyclotron feature at the energy of $E_{\rm cyc} \sim 100 \text{
keV}$ was reported by La Barbera et al. (2001) based on the {\it BeppoSAX}
data.

In the present work, using the data of the {\it NuSTAR} observatory, the
broadband energy spectra, including phase-resolved spectra, of LMC\,X-4 were
investigated at a qualitatively new level for the first time. It allowed us
to obtain limits on the magnetic field in the system and to carry out a
tomography in the fluorescent iron line.

\section*{OBSERVATIONS AND DATA REDUCTION}

\noindent The {\it NuSTAR} (Nuclear Spectroscopic Telescope Array; Harrison
et al., 2013) observatory, launched on June 13, 2012 is a hard X-ray focusing
astronomical telescope, capable to operate in the energy band from 3 to 79
keV. {\it NuSTAR} includes two coaligned telescopes with the focusing systems
and focal plane detector modules (FPMA and FPMB) with an energy resolution of
0.4 keV at 10 keV and 0.9 keV at 60 keV (Harrison et al. 2013).

In this paper we used the publicly available data of the observation of
LMC~X-4 carried out by {\it NuSTAR} at July 4, 2012 with the exposure of
$\sim 39.9 \text{ ksec}$ (ObsID. 10002008001). The superorbital phase during
the observation was around $\Psi_{\rm sup} \simeq 0$ (Molkov et al. 2015),
where the source has a maximum of the persistent flux. During the observation
there were no orbital eclipses (orbital phases $\Psi_{\rm orb} \simeq [0.4;
0.9]$) or X-ray flares. Source events were extracted using a circular region
with the aperture of 120\arcsec, centered at the position of the source ($RA
= 83.206$\arcdeg, $Dec = -66.370$\arcdeg). Background events were extracted
using a polygonal region of the equivalent area. Events were extracted
separately for each of the observatory modules. In order to obtain the better
statistics, the source light-curves, extracted from modules, were combined
following the procedure described in Krivonos et al. (2015).

The primary data processing was carried out using the standard {\it NuSTAR}
data processing software (NuSTAR Data Analysis Software, {\sc nustardas}
version 1.4.1) and the calibration database {\sc caldb} (version 20150612).
Further processing and analysis was carried out using {\sc heasoft} (version
6.17) tools.

A correction of the photon arrival times for the Solar system barycenter was
made using the standard {\sc nustardas} tools. A corresponding correction of
the photon arrival times for the neutron star motion in the binary system was
made out using the orbital parameters from Molkov et al. (2015). Search for
the pulse period was carried out using the epoch folding method (the {\sc
efsearch} tool of the {\sc xspec} package). Pulse profiles were obtained by a
convolution of the source light curves with the measured pulse period. The
spectral analysis was done using the {\sc xspec} package, version~12.8.

\vspace{5mm}

\section*{RESULTS}

\subsection*{Period and pulse profile}

\begin{figure}
\centering
\includegraphics[width=0.47\textwidth]{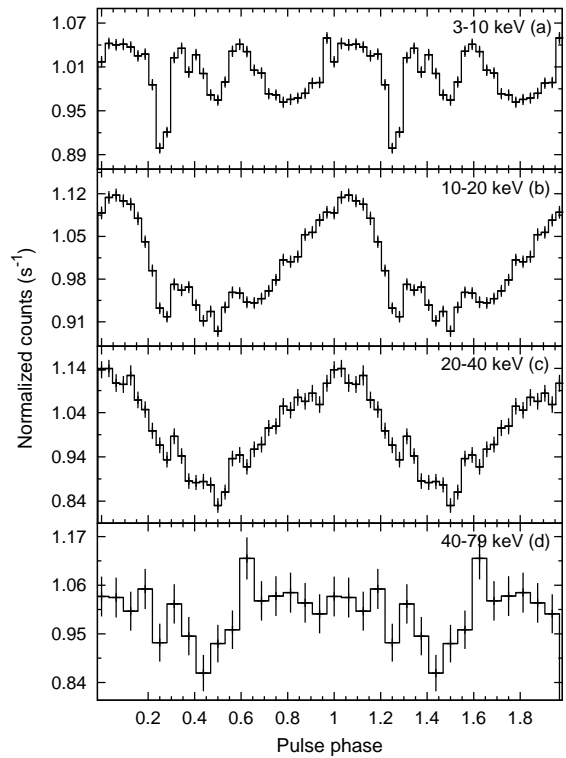}
\caption{Pulse profile of LMC\,X-4 in different energy bands: 3-10~keV~(a),
10-20~keV~(b), 20-40~keV~(c) и 40-79~keV~(d), normalized to the corresponding
average count rate.}
\label{fig:pprof}
\end{figure}

\noindent As it was mentioned above, to determine the pulse period and its
uncertainty we used a combined light curve from both FPMA and FPMB modules.
Using the original light-curve we generated $10^4$ trial light curves (using
the statistics from the original one). Next, for each of the ``test''
light-curves the pulse period value was determined by the epoch folding
technique. The resulting distribution of period values follows a normal
distribution, approximating which by the Gaussian model gives the most
probable value of the pulse period and its error at $1\sigma$ (for details on
the method used see Boldin et al. 2013). As a result we obtained the spin
period of the neutron star at the moment of the {\it NuSTAR} observation
$P_{\rm spin} = 13.49892 \pm 0.00003 \text{ s}$. This value was used for the
further analysis.

The pulse profile contains an important information about the geometry and
the physical properties of emitting regions of a binary system. Figure
\ref{fig:pprof} shows the pulse profiles of LMC~X-4 in four different energy
bands. From figure it can be clearly seen that the pulse profile in two
middle energy bands 10-20 keV and 20-40 keV has a simple single-peaked shape
close to the sinusoidal one, while the pulse profile in a soft energy band
3-10 keV shows a complex structure with several pronounced features. In
particular, there are two additional peaks at the phases $\phi \simeq 0.4$
and $\phi \simeq 0.6$. A similar significant complication of the pulse
profile in soft X-rays has been observed previously (see, e.g., Levine et al.
1991; Woo et al. 1996). It was interpreted as a presence of different
spectral components of the emission with different beam patterns. The signal
above 40 keV is quite weak due to the quick fall of the source intensity with
the increase of the energy and a lack of statistics.

\begin{figure}
\centering
\includegraphics[width=0.55\textwidth]{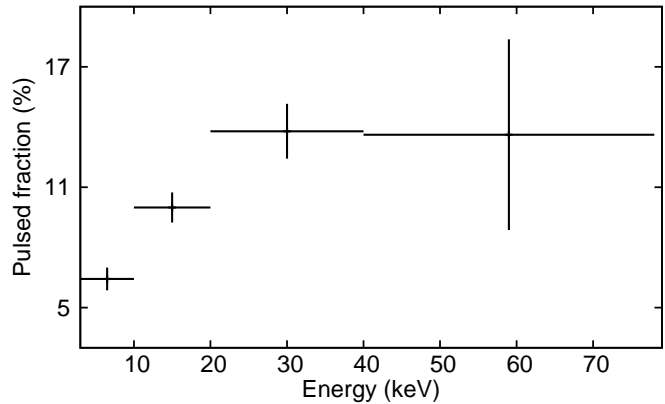}
\caption{Dependence of the pulsed fraction of LMC~X-4 on the energy band.}
\label{fig:pf}
\end{figure}

Figure\,\ref{fig:pf} shows the dependence of the pulsed fraction $PF =
(I_{\rm max} - I_{\rm min})/(I_{\rm max} + I_{\rm min})$ on the energy, where
$I_{\rm max}$ and $I_{\rm min}$ -- maximum and minimum intensities of the
pulse profile in the corresponding energy range. The pulsed fraction of
LMC~X-4 appears to be relatively small: it is just $PF \simeq 6 \%$ in the
energy range 3-10 keV and subsequently increases up to $PF \simeq 14 \%$ in
the energy range 20-40 keV and above. Note, that such a behaviour is typical
for X-ray pulsars, particularly for the bright ones (Lutovinov, Tsygankov 2009).

\begin{figure}
\centering
\includegraphics[width=0.5\textwidth]{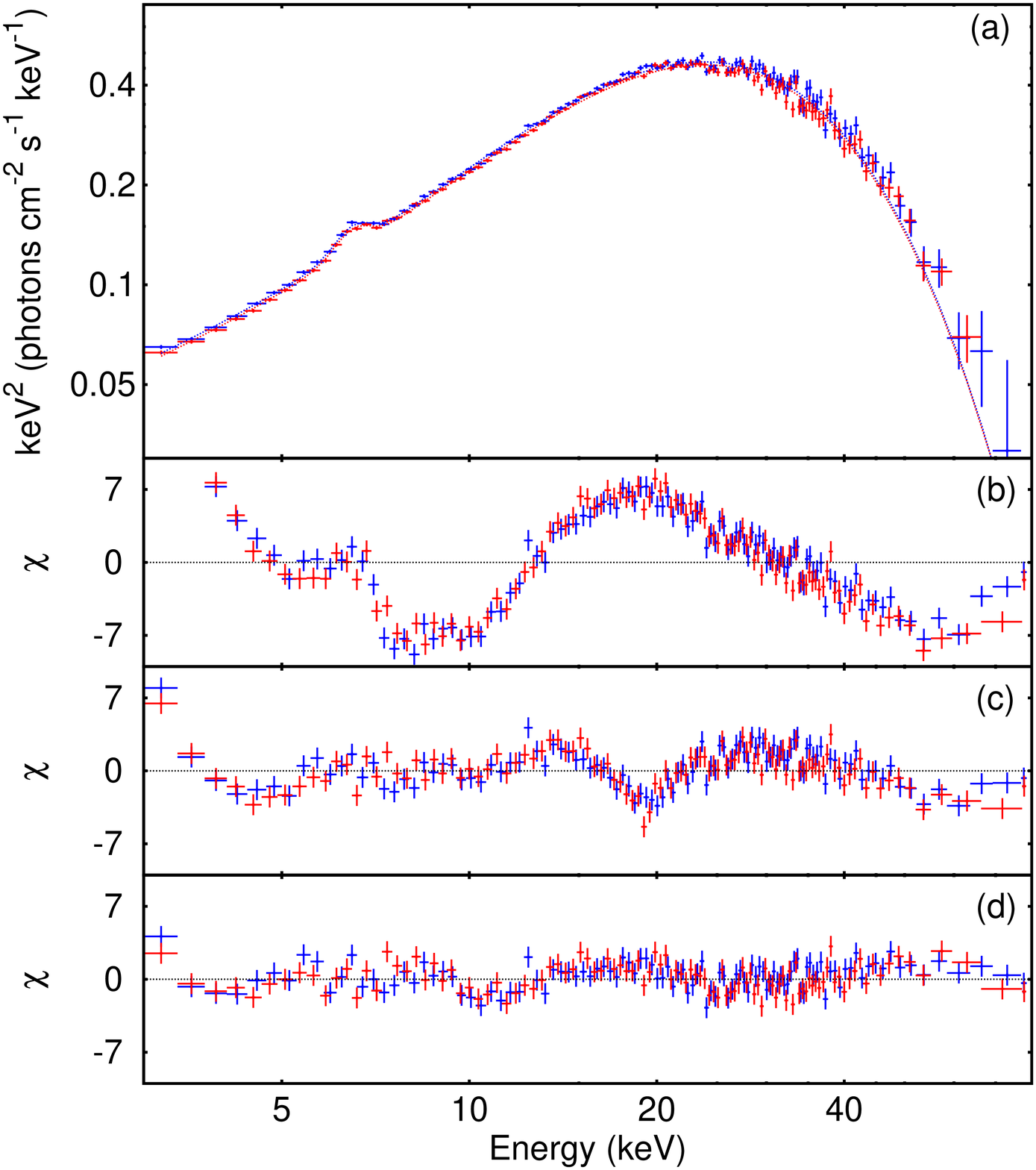}
\caption{Averaged energy spectrum of LMC~X-4 and its best-fit approximation
with the model III (a); deviations of the data from different approximation
models: I (b), II (c) и III (d). Blue and magenta points correspond to the
FPMA and FPMB modules, respectively.}
\label{fig:espec}
\end{figure}

\subsection*{Phase-averaged energy spectra}

\noindent To approximate the spectrum of LMC~X-4 we used several standard
{\sc xspec} models, commonly used to describe spectra of X-ray pulsars: (I)
the power-law model with the exponential cutoff {\sc cutoffpl}, (II) the
power-law model with the high energy cutoff {\sc powerlaw} $\times$ {\sc
highecut} (White et al. 1983) and (III) comptonization model {\sc comptt}
(Titarchuk 1994; Lyubarskii, Titarchuk 1995). To improve the quality of the
approximation a component representing $K_{\alpha}$ iron line was added to
each of models in the form of Gaussian ({\sc gaus xspec} model). The
interstellar absorption was also taken into account by adding the {\sc wabs}
component to the models. Thus, the final models used for the approximation
of the LMC\,X-4 spectra can be written as: 

\noindent
Model I: {\sc wabs(cutoffpl + gaus)}

\noindent
Model II: {\sc wabs(powerlaw $\times$ highecut + gaus)}

\noindent
Model III: {\sc wabs(comptt + gaus)}

Spectra for FPMA and FPMB modules data were analyzed simultaneously. In order
to account the difference between the modules calibration a cross-calibration
constant $C$ between them was included in all spectral models. The $\chi^2$
statistic was used to estimate the approximation quality. All models appeared to
be insensitive to the value of the absorption, therefore in following it
was fixed at $N_{\rm H} = 5.74 \times 10^{20} \text{ atoms cm}^{-2}$
(assuming Solar abundance). A gravitational redshift has been fixed for the
model {\sc comptt} on the value of $z = (1 - 2GM_{\ast}/R_{\ast}c^2)^{-1/2} -
1 = 0.3657$, estimated for the following parameters of the neutron star:
$M_{\ast} = 1.57 M_{\sun}$ and $R_{\ast} = 10^6 \text{ cm}$. Best fit
parameters for models I, II and III are presented in Table 1. The value of
the a normalization coefficient was $C=1.021\pm0.004$ for all the
models.

\begin{table*}
\centering
\caption{Parameters of the best approximation of the spectrum models of the LMC~X-4}
	\begin{tabular}{cccc}
		\hline\hline
		Parameter / Model & Model I & Model II & Model III \\
		\hline
		$N_{\rm H} (\times 10^{20} \text{ cm}^{-2})$  & 5.74 & 5.74 & 5.74 \\
		$T_{0} \text{ (keV)}$ & ... & ... & 0.56 $\pm$ 0.08 \\
		$kT \text{ (keV)}$ & ... & ... & 9.08 $\pm$ 0.04 \\
		$\tau$ & ... & ... & 14.15 $\pm$ 0.07 \\
		$\Gamma$ & 0.21 $\pm$ 0.01 & 0.834 $\pm$ 0.004 & ... \\
		$E_{\rm cut} \text{ (keV)}$ & ... & 19.35 $\pm$ 0.16 & ... \\
		$E_{\rm fold} \text{ (keV)}$ & 14.11 $\pm$ 0.13 & 14.98 $\pm$ 0.18 & ... \\
		$E_{\rm Fe} \text{ (keV)}$ & 6.43 $\pm$ 0.04 & 6.48 $\pm$ 0.03 & 6.46 $\pm$ 0.03 \\
		$\sigma_{\rm Fe} \text{ (keV)}$ & 0.11 $\pm$ 0.05 & 0.26 $\pm$ 0.03 & 0.41 $\pm$ 0.05 \\
		$EW_{\rm Fe} \text{ (eV)}$ & 41 $\pm$ 8  & 95 $\pm$ 7 & 158 $\pm$ 3 \\[2mm]
		
		$\chi^2$ / d.o.f. & 3.47 (2054) & 1.27 (2053) & 1.10 (2053) \\
		\hline
	\end{tabular}
\end{table*}

Figure\,\ref{fig:espec}a shows the phase-averaged energy spectrum of LMC~X-4
and its best-fit approximation with the model~III. Deviations of the data
from the approximation models I, II and III are shown in panels (b), (c) and
(d), respectively. Both from the Table 1 and Figure\,\ref{fig:espec} it can
be clearly seen that the model~III gives the best approximation with the
value of $\chi^2 = 1.10$ for a 2053 degrees of freedom (d.o.f). The source
flux in the energy range 3-79 keV is of $F_{\rm x} = (1.20 \pm 0.05) \times
10^{-9} \text{ erg cm}^{-2} \text{ s}^{-1}$.

\subsection*{Phase-resolved spectroscopy}

\noindent Pulse phase-resolved spectra of LMC~X-4 were accounted in 16
uniformly distributed phase bins and approximated with the comptonization model
III. Event list for each phase bin was created by selecting events in
corresponding time intervals. This procedure was applied both for FPMA and
FPMB data. We used the same approaches as for the phase-averaged
spectra analysis for the approximation of the pulse phase-resolved spectra
and estimate the quality of this approximation.

\begin{figure*}
\centering
\includegraphics[width=0.8\textwidth]{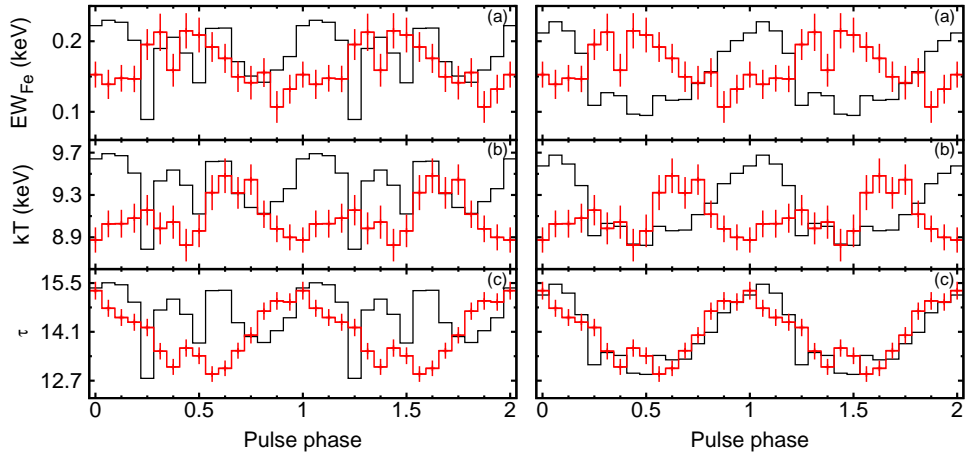}
\caption{Spectral parameters of the model III as a function of the pulse phase
(red histograms) in a comparison with the source pulse profile in the 3-10
(left panel) and 10-20 keV (right panel) energy ranges.}
\label{img:ph}
\end{figure*}

The $\chi^2$\,(d.o.f.) value of the best-fit approximation of the pulse
phase-resolved spectra varies from 0.88 to 1.12 for $\sim1000$ degrees of
freedom, that indicates an acceptable quality of the approximation.
Figure\,\ref{img:ph} shows parameters of the model III as a function of the
pulse phase in a comparison with the source pulse profile in the 3-10 and
10-20 keV energy ranges. It can be clearly seen that the characteristic
temperature and the optical depth of the Compton emission are changed
significantly with the pulse profile. The optical depth correlates with the
pulse profile, while the maximum temperature is somewhat displaced in
relation to the profile maximum.

At the same time, the behavior of the equivalent width of the iron line looks
most interesting and demonstrates a significant phase shift in a comparison
to the emission maximum. Based on the Fig.\,\ref{img:ph} we can estimate this
phase shift as $\Delta \phi \sim 0.4$. Taking into account the source pulse
period $P_{\rm spin} \simeq 13.5 \text{ sec}$ the corresponding time delay is
around $\Delta t \sim 5 \text{ sec}$. This delay is likely to be related with
the travel time of photons between the regions of their emission in the
vicinity of the neutron star and the place where they are reflected. The
distance which can be travelled by photons during $\sim5$ seconds is about
$\sim 1.5 \times 10^{11} \text{ cm}$. Such a distance corresponds roughly to
the outer radius of the accretion disk and is consistent with estimates
presented by Neilsen et al. (2009) from the analysis of the Doppler
broadening of the iron line.

\subsection*{Search for the cyclotron absorption line}

\noindent To check the hypothesis of a possible presence of a cyclotron
absorption line in the spectrum of the X-ray pulsar LMC~X-4, the model III
was modified by addition of the {\sc gabs} component from the {\sc xspec}
package. Following the procedure applied by Tsygankov, Lutovinov (2005), the
cyclotron line energy $E_{\rm cyc}$ was varied within the 5-55 keV energy
range with the step of 3 keV. A corresponding line width $W$ was varied
within 4-8 keV energy range with the step of 2 keV. For each pair of
cyclotron line parameters the energy and the width were fixed in the {\sc
gabs} component and the resulting model was used to approximate the source
spectrum. As a result, none of the combination of the line energy and its
width does not result in a significant improvement of the fit and only
the upper limit for the optical depth can be obtained.

\begin{figure}
\centering
\includegraphics[width=0.5\textwidth]{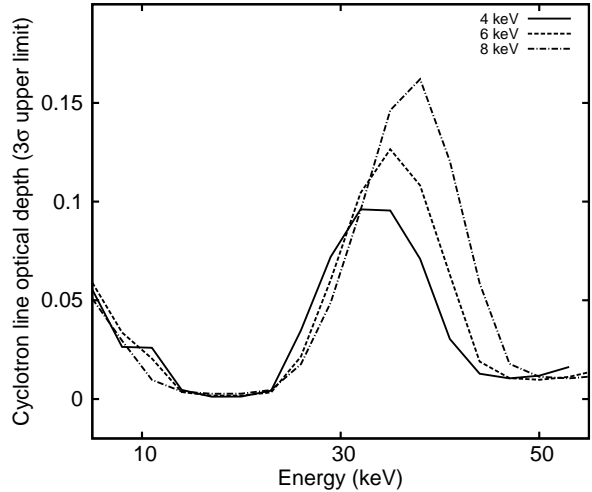}
\centering	
\caption{Dependence of the upper limit ($3\sigma$) on the optical depth of
the cyclotron absorption line on the energy $E_{\rm cyc}$ for three different
line width: 4~keV (solid line), 6~keV (dashed line) and 8~keV (dash-dotted line).}
\label{img:ecyctau}
\end{figure}

Figure\,\ref{img:ecyctau} shows the dependence for the upper limit for the
optical depth of cyclotron line on the energy $E_{\rm cyc}$ for the three
possible line widths: 4, 6 and 8 keV. The maximum value of the upper limit
for the optical depth of the cyclotron line is $\sim 0.15$ ($3\sigma$). A
similar search for the cyclotron absorption line in LMC~X-4 was carried out
for the pulse phase-resolved spectra. The corresponding upper limit on the
optical depth is about $\sim 0.4$ ($3\sigma$).

\subsection*{Power spectrum}

\noindent Typically, the power spectrum of X-ray pulsars consists of the red
noise component plus peaks at the pulsar spin frequency and its harmonics.
According to the perturbation propagation model (Lyubarskii 1997; Churazov
2001), the red noise is generated in the accretion disk as a superposition of
individual components arising at each specific radius at the corresponding
characteristic diffuse timescale. The noise propagation inwards to the inner
edge of the disk lead to the corresponding modulation of the accretion rate
onto the compact object. The resulting power spectral density (PSD) has a
form of a self-similar power-law up to the maximum frequency which can be
generated in the disk (Hoshino, Takeshima 1993; Revnivtsev et al. 2009). In
particular, Revnivtsev et al. (2009) showed that the power spectra of X-ray
pulsars in the corotation contain a break at the frequency equal to the spin
frequency of the neutron star (or in another words to the Keplerian
frequency; $f_{\rm br} \approx \nu_{\rm k}$).

\begin{figure}
\centering
\includegraphics[width=0.6\textwidth]{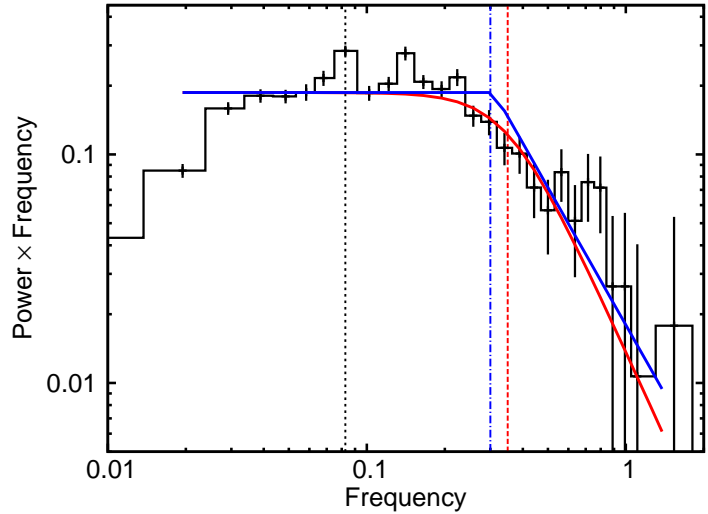}
\caption{Power spectral density of LMC\,X-4. Solid blue line represents the
spectrum approximation with the model 1 ($\alpha_2 = -2.0$, $f_{\rm br} =
0.3$\,Hz). Solid red line represents spectrum approximation with the model 2
($\alpha = -2.5$, $f_{\rm br} = 0.35$\,Hz). Vertical lines show positions of
the corresponding break frequency: $f_{\rm br} = 0.3 \text{ Hz}$ (dash-dotted
line) and $f_{\rm br} = 0.35 \text{ Hz}$ (dashed line); and the position of 
the spin frequency (dotted line).}
\label{fig:powspec}
\end{figure}

Figure\,\ref{fig:powspec} shows the power spectrum of LMC~X-4 in the energy
range 3-79 keV in units of the power multiplied by the frequency. Two models
were used for the approximation:\\
a broken power-law model, the model 1:
\begin{equation} \label{eq:psan1}
P \propto 
\begin{cases}
f^{\alpha_1}, [f < f_{\rm br}]\\
f_{br}^{\alpha_1 - \alpha_2} f^{\alpha_2}, [f > f_{\rm br}]
\end{cases},
\end{equation}

\noindent
where $\alpha_1$ и $\alpha_2$ --  slopes of the
power-law, $\alpha_1$ value was fixed at -1, $f_{\rm br}$ - break frequency;\\

and a model with a smooth transition between the two power-law functions
(Revnivtsev et al. 2009), the model 2:

\begin{equation} \label{eq:psan2}
P \propto f^{-1} \times \left[1 + \left(\dfrac{f}{f_{\rm br}}\right)^4\right]^{-\alpha/4},
\end{equation}

\noindent
where $f_{\rm br}$ -- break frequency and $\alpha$
-- the power-law index.

Results of the PSD approximation with both models are shown in
Fig.~\ref{fig:powspec}. The solid blue line corresponds to the model 1 with
the following parameters: the break frequency $f_{\rm br} \simeq 0.3 \text{
Hz}$, the power-law index $\alpha_2 = -2.0$. The solid red line shows the
result of  the approximation by the model 2 with the following parameters:
the break frequency $f_{\rm br} \simeq 0.35 \text{ Hz}$, the power-law index
$\alpha = -2.5$.


\subsection*{X-ray tomography of the emitting regions}

\noindent The {\it NuSTAR} observation covers approximately a half of the
orbital cycle of the binary system ($\Psi_{\rm orb} \simeq [0.4;0.9]$). It
makes possible to carry out a tomography of the areas where X-ray emission is
generated. In particular, if the iron line Fe\,$\rm K_{\alpha}$ is generated
due to reflection of the original emission in certain areas (regions) of the
binary system, then observing the system at different angles (in different
orbital phases), one can attempt to locate these regions and their geometry.
The energy resolution of the {\it NuSTAR} observatory doesn't allow to carry
out a detailed Doppler imaging of the object. Therefore we confine ourselves
by a rough approximation, which allow us to get qualitative estimates on
time delay in the system.

Using the {\it Chandra} and {\it XMM-Newton} observations Neilsen et al.
(2009) carried out the analysis of regions of the formation of the lines in
the spectrum of LMC~X-4. These authors found that changes in the intensity
and width of spectral lines of different elements depend on the phase of
superorbital motion. They also proposed three possible areas of the lines
generation in the binary system: the photoionized stellar wind region, the
outer region of the standard accretion disk and the inner region of the
curved accretion disk. According to the assumptions of Neilsen et al. (2009)
the most probable area of the formation of the Fe\,$\rm K_{\alpha}$ line is
the outer edge of the accretion disk and/or the inflow accretion stream,
which falls on the outer edge of the disk through the inner Lagrange point.
If the Fe\,$\rm K_{\alpha}$ line is generated due to the irradiation of the
pulsar emission at the outer edge of the accretion disk or at the inflow
stream, one can expect a correlation between the pulse profile and the line
parameters (e.g., equivalent width). Moreover, this correlation will look
different for different orbital phases, depending on the place where the line
is formed -- at the outer edge of the accretion disk or in the place where
inflow accretion stream touches the outer edge of the disk (the so-called hot
spot, see Armitrage, Livio 1996 and references therein).

\begin{table*}
	\centering
	\captionof{table}{Parameters of the LMC\,X-4 spectra in different orbital phases.}
	\begin{tabular}{cccc}
		\hline\hline
		Parameter / Phase & $\Psi_{\rm orb}^{(1)}$ & $\Psi_{\rm orb}^{(2)}$ & $\Psi_{\rm orb}^{(3)}$ \\
		\hline
		$C$ & $1.021 \pm 0.006$ & $1.016 \pm 0.006$ & $1.034 \pm 0.006$ \\
		$N_{\rm H} (\times 10^{20} \text{ cm}^{-2})$  & 5.74 & 5.74 & 5.74 \\
		$T_{0} \text{ (keV)}$ & 0.56 & 0.56 & 0.56 \\
		$kT \text{ (keV)}$ & 9.03 $\pm$ 0.06 & 9.10 $\pm$ 0.06 & 8.99 $\pm$ 0.06 \\
		$\tau$ & 13.91 $\pm$ 0.10 & 13.94 $\pm$ 0.10 & 14.3 $\pm$ 0.11 \\
		$E_{\rm Fe} \text{ (keV)}$ & 6.49 $\pm$ 0.05 & 6.53 $\pm$ 0.06 & 6.38 $\pm$ 0.07 \\
		$\sigma_{\rm Fe} \text{ (keV)}$ & 0.34 $\pm$ 0.09 & 0.35 $\pm$ 0.07 & 0.54 $\pm$ 0.09 \\
		$EW_{\rm Fe} \text{ (eV)}$ & 140 $\pm$ 10  & 142 $\pm$ 11 & 218 $\pm$ 15 \\[2mm]

		$\chi^2$ / d.o.f. & 1.05 (1531) & 1.01 (1528) & 1.07 (1529) \\
		\hline
	\end{tabular}
\end{table*}

Indeed, for the superorbital phase $\Psi_{\rm sup} \sim 0$ the plane of the
accretion disk is positioned in the direction to the observer at an angle
close to the normal. If the iron line is generated at the outer edge of the
accretion disk, we will observe a constant phase shift between the maximum of
its equivalent width and the maximum of the pulse profile during the binary
motion of the neutron star. If the Fe\,$\rm K_{\alpha}$ line is produced in
the inflow stream (hot spot), one can expect variations in the phase shift or
in the width profile, depending on the orbital phase. This is because the
system asymmetry in relation to the distant observer, defined by the presence
of the hot spot or the inflow stream on the disk edge. To check this
assumptions, the entire observation was divided into three equal subintervals
(orbital phases: $\Psi_{\rm orb}^{(1)} \simeq [0.40-0.56]$, $\Psi_{\rm
orb}^{(2)} \simeq [0.56-0.72]$ и $\Psi_{\rm orb}^{(3)} \simeq [0.72-0.90]$).
Both phase-averaged and pulse phase-resolved spectra were produced in every
subinterval. As before, they all were approximated by the comptonization
model III.

The best fit parameters of the pulse-averaged spectra for each of the three
orbital phases are presented in Table~2. The temperature of the seed photons
$T_{0}$ is restricted poorly due to lack of statistics, therefore it was
fixed at $T_{0} = 0.56 \text{ keV}$, measured for the total spectrum (see
Table~1). It can be seen from the Table~2 that for all the three orbital
phases the spectral parameters coincide to each other within the
uncertainties, except the iron line equivalent width. In the third orbital
phase interval ($\Psi_{\rm orb}^{(3)}$) the equivalent width of the line of
iron ($EW_{\rm Fe}$) by factor of $\sim 1.5$ higher than in the first and
second phase intervals.

\begin{figure*}
\includegraphics[width=\textwidth]{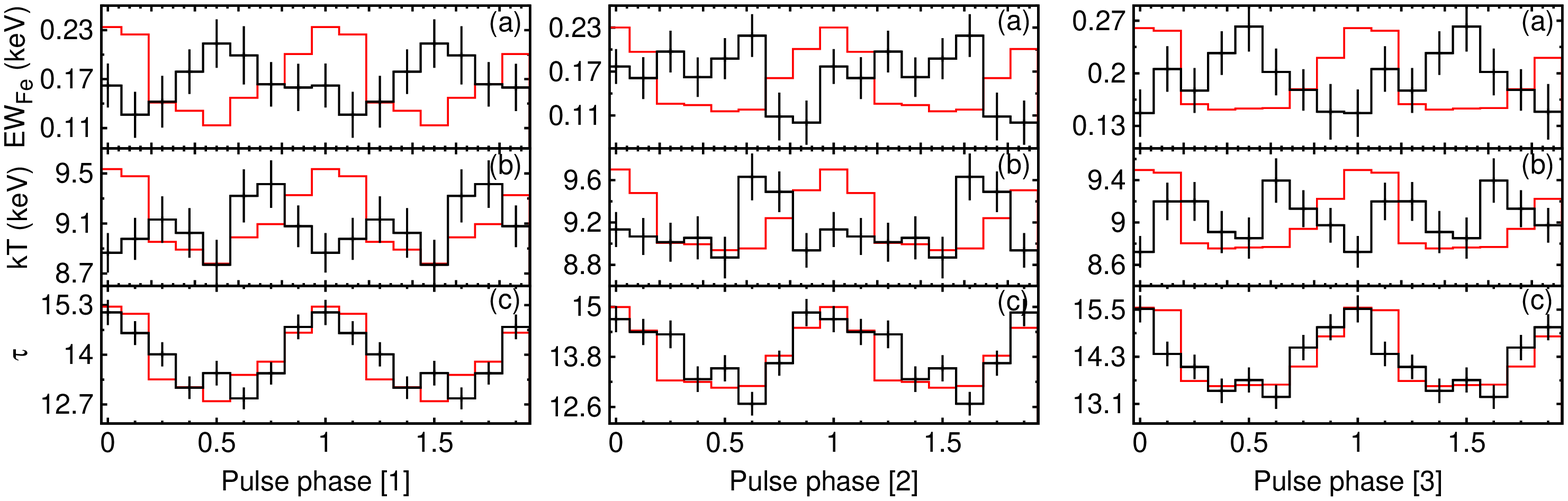}
\caption{Dependence of the spectral parameters of LMC\,X-4 on the pulse phase
for different orbital phases $\Psi_{\rm orb}^{(1)}$, $\Psi_{\rm orb}^{(2)}$
and $\Psi_{\rm orb}^{(3)}$ (red histograms). The black line in each panel
shows the corresponding pulse profile in a 10-20 keV energy range. }
\label{fig:pfitall}
\end{figure*}

\begin{figure*}
\centering
\includegraphics[width=0.9\textwidth]{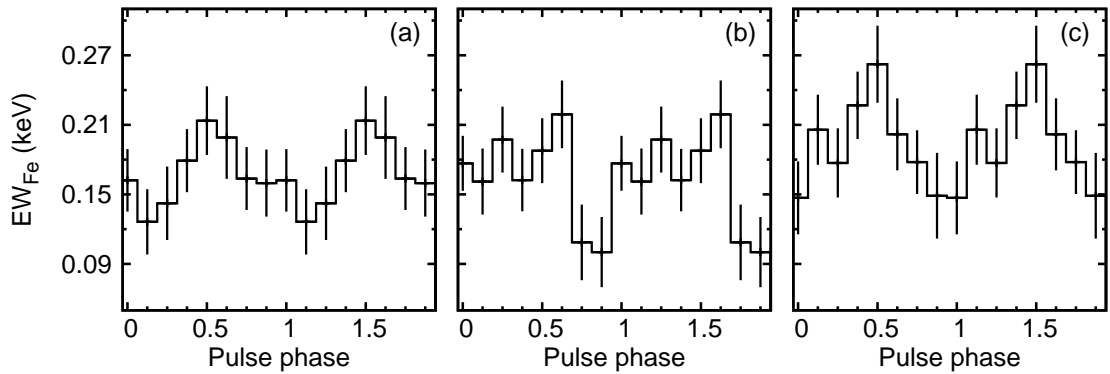}
\caption{Dependence of the equivalent width of the iron line on the pulse
phase in different orbital phases $\Psi_{\rm orb}^{(1)}$ (a),
$\Psi_{\rm orb}^{(2)}$ (b) and $\Psi_{\rm orb}^{(3)}$ (c). See text for
details.}
\label{fig:wefeall}
\end{figure*}

Figure\,\ref{fig:pfitall} shows variations of the spectral parameters with
the pulse phase for each of the three orbital intervals in a comparison with
corresponding source pulse profile in the 10-20~keV energy range. One can be
seen that at orbital phases $\Psi_{\rm orb}^{(1)}$ and $\Psi_{\rm or}^{(3)}$
an anti-correlation between the pulse profile and the equivalent width of
Fe\,$\rm K_{\alpha}$ is observed. Moreover, the shape of the equivalent width
changes at these phases has a pronounced maximum and minimum shifted about
the half of the pulse phase. At the same time at the orbital phases
$\Psi_{\rm orb}^{(2)}$ the shape of the equivalent width becomes asymmetric
and its maximum shifts by $\sim0.1$ of the pulse phase. Thus, results of the
pulse phase-resolved spectroscopy demonstrates clearly variations in the
phase shift and in the width profile depending on the orbital motion of the
pulsar. For the better clarity all the three profiles of the equivalent line
width Fe\,$\rm K_{\alpha}$ is presented in-line in Fig.~\ref{fig:wefeall}.

Based on these results one can assume that the generation of the Fe\,$\rm
K_{\alpha}$ line occurs in the inflowing stream or in the area of the
interaction between the stream and the outer edge of the accretion disk. It
should be noticed that the interaction between the stream and the accretion
disk is quite complex and depends on a large number of parameters of the
binary system and peculiarities of the matter transfer between the components
(see, e.g., Bisikalo et al. 2013). In particular, the hot spot itself doesn't
always appear in the place of the interaction of the inflowing stream and the
accretion disk. Therefore, additional observations, involving data from other
observatories, are needed for the detailed analysis and obtaining the
quantitative parameters of the Fe\,$\rm K_{\alpha}$ line generation area.

\section*{CONCLUSION}

\noindent In this paper, using the {\it NuSTAR} observatory data, the
broadband spectrum of the X-ray pulsar LMC~X-4 was obtained in the energy
range $3 - 79 \text{ keV}$ with the high statistical significance and good
energy resolution. It allowed us to test several spectral models and to
determine their parameters. It was shown that the obtained spectrum can be
approximated in the best way by the comptonization model ({\sc comptt}) with
the inclusion of the interstellar absorption and the $K_{\alpha}$ iron line.
The equivalent width of the iron line in the averaged spectrum was of
$EW_{\rm Fe} \simeq 158 \text{ eV}$.

We searched for the cyclotron absorption line in the source averaged spectrum
in the energy range $5 - 55 \text{ keV}$ and for different possible
equivalent line width (2-8~keV). The resulting upper limit on its optical
depth $\simeq 0.15$ ($3\sigma$) indicates the absence of the cyclotron
feature in a given energy range. This allows us to put a limit on the
possible magnitude of the magnetic field on the surface of the neutron star
in LMC~X-4: $B < 3 \times 10^{11}\text{ G}$ or $B > 6.5\times 10^{12} \text{
G}$. Search for the cyclotron absorption line was also carried out for
pulse phase-resolved spectra and also showed the absence of the cyclotron
feature in the energy range mentioned above.

The pulse phase-resolved spectroscopy was carried out for LMC\,X-4 for the
first time in the wide energy band with the good spectral resolution. The
phase-resolved spectra were approximated by the same comptonization model as
the averaged one. The comparison of the behavior of the spectral parameters
with the pulse profile at the energies above 10~keV has shown a correlation
of the source intensity with the Compton optical depth and anti-correlation
with the temperature and the equivalent width of the iron line. We determined
the delay ($\sim 5 \text{ sec}$) between the peaks of the source emission and
the equivalent width of the iron line, apparently associated with the travel
time of photons between the emitting regions in the vicinity of the neutron
star and the relatively cold regions where this emission is reflected
(presumably, at the inflowing stream or at the place of an interaction of the
stream and the outer edge of the accretion disk). The tomography analysis of
the Fe\,$\rm K_{\alpha}$ line implies that this emission originates not just
on the outer edge of the accretion disk, but most likely in the inflowing
accretion stream or in the area where this flow interacts with the outer edge
of the accretion disk (the so-called hot spot). However this assumption
requires more detailed study. It is also worth noticing that in the energy
range of 3--10~keV, where the pulse profile has a more complex shape, there
were no clear correlation between the model parameters and the observed pulse
profile.

The tomography of the emitting areas allows us to suggest that the generation
of the Fe\,$\rm K_{\alpha}$ line, emerging as a result of the reflection of
the original X-ray flux in relatively cold areas of the surrounding matter,
is not just on the outer edge of the accretion disk, but most likely in the
inflowing stream that flows through inner Lagrangian point and falls on the
outer accretion disk edge or in the area where flow interacts outer edge of
the accretion disk (the so-called hot spot).

However this assumption requires more detailed study. It is also worth
noticing that in in energy range of 3--10~keV, where the pulse profile has a
more complex shape, there were no clear correlation between the model
parameters and the form of profile of the pulse observed.

Revnivtsev et al. (2009) and Tsygankov et al. (2012) demonstrated the
possibility of using the noise power spectral density shape to determine the
magnetic field strength of the neutron star in X-ray pulsars. This method is
based on the equality of the break frequency in the PSD and the Keplerian
frequency on the inner edge of the accretion disk ($f_{\rm br} \approx
\nu_{\rm k}$; Revnivtsev et al 2009).

Substituting in the expression for the Keplerian motion the mass of the
neutron star in LMC~X-4 $M_{\ast} = 1.57 M_{\sun}$ and the break frequency,
measured from the PSD $f_{\rm br} \simeq (0.30-0.35) \text{ Hz}$, we can
estimate the magnetospheric radius as $R_{\rm m} \sim (3.5 - 3.9) \times
10^{8} \text{ cm}$. The radius of the magnetosphere, in turn, is determined by
the strength of the magnetic field of the neutron star (see, e.g., Tsygankov
et al., 2016):

\be
R_{\rm m} =\xi \dot{M}^{-2/7} \mu^{4/7}  (2GM)^{-1/7} ,
\ee

\noindent
where $\xi$ -- dimensionless coefficient, taking into account non-spherical
nature of accretion ($\xi \sim 0.5$; Long et al. 2005; Parfrey et al. 2015),
$\mu$ -- magnetic moment of the neutron star and $\dot{M}$ -- mass accretion
rate.

The flux measured from the pulsar in the 3-79~keV energy band is $F_{\rm x}
\sim 1.20 \times 10^{-9} \text{erg cm}^{-2} \text{sec}^{-1}$ which can be
roughly considered as a bolometric one (the flux in the 0.5-100~keV energy
range is $F_{\rm x} \sim 1.22 \times 10^{-9} \text{erg cm}^{-2}
\text{sec}^{-1}$). This corresponds to the bolometric luminosity of the
source $L_{\rm x} \sim 3.6 \times 10^{38} \text{ erg sec}^{-1}$ and the mass
accretion rate $\dot{M} \sim 6.3 \times 10^{-8} M_{\sun} \text{ year}^{-1}$.
By making the Keplerian radius equal to the magnetospheric radius we can
estimate the magnetic field strength of the neutron star in LMC~X-4: $B
\simeq (2.7-3.2) \times 10^{13} \text{ G}$, which can be roughly translated
to the energy of the possible cyclotron line $230-270 \text{ keV}$. Formally,
this value is agreed with the lower limit obtained from the analysis of the
energy spectrum of the pulsar.

\section*{Acknowledgements}

This work was supported by grant RNF~14-12-01287. The research has made use
of data obtained with {\it NuSTAR}, a project led by Caltech, funded by NASA
and managed by NASA/JPL, and has utilized the NUSTARDAS software package,
jointly developed by the ASDC (Italy) and Caltech (USA).  The authors thanks
to R.A.Krivonos for the help with the processing of the {\it NuSTAR}
observatory data.

\end{document}